\pdfoutput=1
\documentclass[a4paper,fleqn,usenatbib]{mnras}

\usepackage[T1]{fontenc}
\usepackage{ae,aecompl}
\usepackage{graphicx}
\usepackage{physics}
\usepackage{amsmath}

\usepackage{graphicx}	
\usepackage{amsmath}	
\usepackage{amssymb}	
\newcommand{\msun}{{\rm M}_\odot}
\usepackage{newtxtext,newtxmath}

\usepackage[normalem]{ulem}
\usepackage{multirow}
\newlength{\width}
\title[Pulsations of primordial supermassive stars.]{Pulsations of primordial supermassive stars induced by a general relativistic instability; visible to JWST at z$>$12. }

\author[C. Nagele et al.]{
Chris Nagele,$^{1}$\thanks{E-mail: chrisnagele.astro@gmail.com}
Hideyuki Umeda,$^{1}$
Koh Takahashi,$^{2}$
Keiichi Maeda$^{3}$
\\
$^{1}$Department of Astronomy, Graduate School of Science, the University of Tokyo, Tokyo, 113-0033, Japan\\
$^{2}$Astronomical Institute, Graduate School of Science, Tohoku University, Sendai, 980-8578, Japan\\
$^{3}$Department of Astronomy, Kyoto University, Kitashirakawa-Oiwake-cho, Sakyo-ku, Kyoto 606-
8502\\
}

\date{Accepted XXX. Received YYY; in original form ZZZ}

\pubyear{2022}

\begin{document}
\label{firstpage}
\pagerange{\pageref{firstpage}--\pageref{lastpage}}
\maketitle

\begin{abstract}
The origin of high-redshift quasars and their supermassive black hole engines is unclear. One promising solution is the collapse of a primordial supermassive star. Observational confirmation of this scenario may be challenging, but a general relativistic instability supernova provides one avenue for such. Previous studies have found that a general relativistic instability supernova has a potentially decades-long plateau phase visible to JWST at high redshift. In this work, we examine stars with mass just below the general relativistic instability supernova mass range. These stars pulsate, ejecting a portion of their envelopes. They then contract quasi-statically back to an equilibrium temperature, at which point they again become unstable and pulsate once more. Because each pulse consumes a small amount of the available nuclear fuel, there exists the possibility of multiple pulsations. We present simulations of the contracting phase, the pulsation, and the light-curve phase. We find that the lower mass pulsating models are even brighter than the higher mass supernovae because the pulsations occur in the late helium burning phase when the stars have extremely large radii. The fact that the pulsations are more luminous and occur in a wider mass range than the supernovae bodes well for observation.

\end{abstract}



\suppressfloats

\section{Introduction}
\label{introduction}

The first data release of the JWST has allowed us to peer further back in time than ever before. After mere weeks of observation, JWST has reportedly detected an unexpectedly large population of galaxies above redshift ten \citep{Adams2022arXiv220711217A,Atek2022arXiv220712338A,Harikane2022arXiv220801612H}, with several candidates even at z $\approx$ 20 \citep{Yan2022arXiv220711558Y}. A few of these high-redshift galaxies are more luminous than anticipated \citep{Boylan-Kolchin2022arXiv220801611B}, a fact which, if confirmed \citep[e.g.][]{Naidu2022arXiv220802794N}, would require a reassessment of our understanding of galaxy formation in the early universe. JWST has also enabled metallicity measurements at higher redshift than ever before \citep{Curti2022arXiv220712375C} and is expected to greatly improve our understanding of supermassive black holes (SMBHs) throughout cosmic history \citep{gardner2006}.

For some time, the community has known that the study of SMBHs and the study of the early universe must overlap. Quasars powered by enormous SMBHs have been observed as early as redshift seven \citep{mortlock2011,wu2015,banados2018,matsuoka2019,wang2021}. The origin of these objects is unknown, as no simple model can explain their existence \citep{inayoshi2020}. This has led to the consideration of more complicated models, such as the direct collapse scenario, whereby a primordial supermassive star (SMS, M $>10^3\;\msun$) is formed and, at the end of its relatively short life, it becomes a seed for the observed SMBHs \citep{bromm2003,Volonteri2010MNRAS.409.1022V,Woods2019PASA...36...27W,Haemmerle2020SSRv..216...48H}. 

The detection of these SMSs by JWST has been suggested, either directly \citep{Zackrisson2010MNRAS.407L..74Z,surace2018,surace2019,Martins2020A&A...633A...9M,Woods2021ApJ...920L..22W,Vikaeus2022ApJ...933L...8V} or indirectly \citep{whalen2013,moriya2021} via a general relativistic instability supernova (GRSN) \citep{chen2014,nagele2020,Nagele2022arXiv220510493N}. The GRSN is the confluence of the explosive alpha process and the general relativistic (GR) radial instability. The alpha process is extremely efficient in terms of energy per unit mass, so unusual circumstances are required for it to power a supernova; circumstances such as a $\sim 10^4\;\msun$ star which is destabilized by the effects of GR.

In this letter, we introduce another--- somewhat fortuitous--- element to the above picture. In our previous paper \citep{Nagele2022arXiv220510493N}, we reported the discovery of pulsations, which are similar to GRSNe, but only eject a portion of the envelope, analogous to the pulsational pair instability (PPI) process \citep[e.g.][]{Yoshida2016MNRAS.457..351Y,woosley2017}. In the PPI process, subsequent pulsations tend to have lower energy because each pulse consumes a significant portion of the star's fuel (at least for high CO core masses, see \citealt{Renzo2020A&A...640A..56R}). However, because the alpha process is so efficient, SMS pulsations consume only a small fraction of the fuel in the SMS core. Thus, if the next GR instability triggers after the core has mixed convectively, then the chemical composition will be effectively unchanged and the SMS will undergo another pulsation. We examine the properties and compute the lightcurves of the first and second pulsations for several models from \citet{Nagele2022arXiv220510493N}. In Sec. \ref{methods}, we describe our numerical methods and procedures, in Sec. \ref{results} we report out findings, and we close with conclusions and a discussion in Sec. \ref{discussion}.

\section{Methods}
\label{methods}

To properly model the post pulsational evolution of the supermassive star, we use three codes. First, we port the final profiles from the calculations of \citet{Nagele2022arXiv220510493N} back into our post Newtonian stellar evolution code, HOSHI. We then follow the quasi-static contraction of the star from its inflated, post pulsational state back to temperatures capable of supporting the star via nuclear burning. Near the end of this contraction, the star once again becomes GR unstable, and we port it back to our GR hydrodynamics code, HYDnuc. In HYDnuc, the star pulsates again, after which we port the star to SNEC in order to calculate the light curve resulting from the pulsation. 

\begin{figure}
    \centering
    \includegraphics[width=\columnwidth]{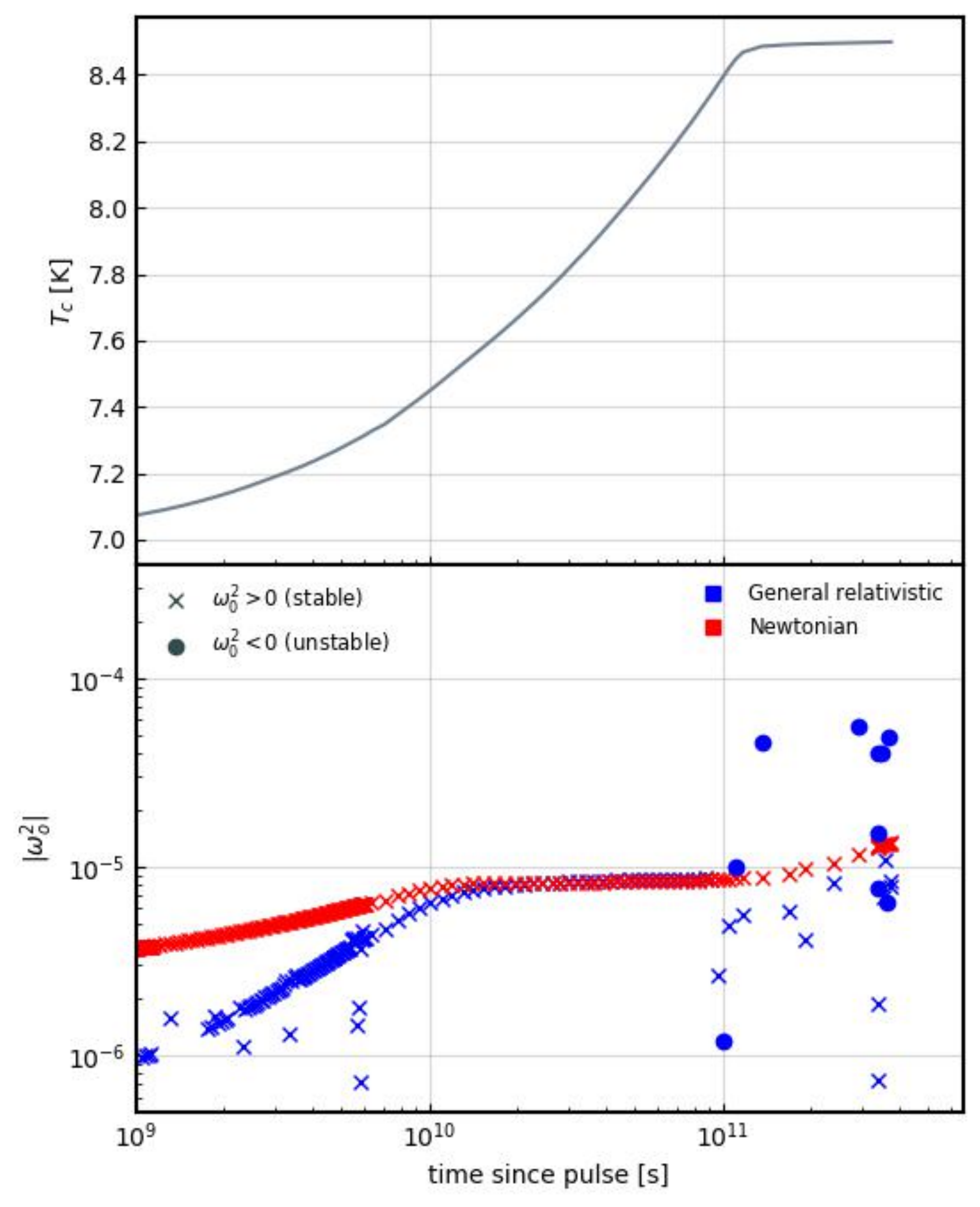}
    \caption{Results of the stellar evolution calculation and stability analysis after the first pulse. Upper panel --- evolution of central temperature in HOSHI. The Kelvin Helmholtz timescale is order $10^{11}$ s. Lower panel --- GR stability analysis. The star becomes unstable near the end of the quasi-static contraction.}
    \label{fig:HOSHI}
\end{figure}

The HOSHI code is a stellar evolution and hydrodynamics code including the first order, static, post-Newtonian (PN) corrections to GR. This code includes all the necessary ingredients to model SMS evolution, particularly a nuclear reaction network (52 isotopes), neutrino cooling, mass loss, rotation (in this paper we consider only slow rotators), radiative and convective energy transport, and an equation of state with contributions from photons, averaged nuclei, electrons, and positrons. We port the initial models by taking the last pre-pulsation stellar model and altering it's temperature, density, and chemical composition to match the post-pulsational HYDnuc model. We evolve the SMS in HOSHI until it becomes unstable (Fig. \ref{fig:HOSHI}) which corresponds to the end of the quasi-static contraction phase, the timescale of which is on the same order of magnitude as the Kelvin Helmholtz timescale of the star ($10^{11}$ s).

\begin{figure}
    \centering
    \includegraphics[width=\columnwidth]{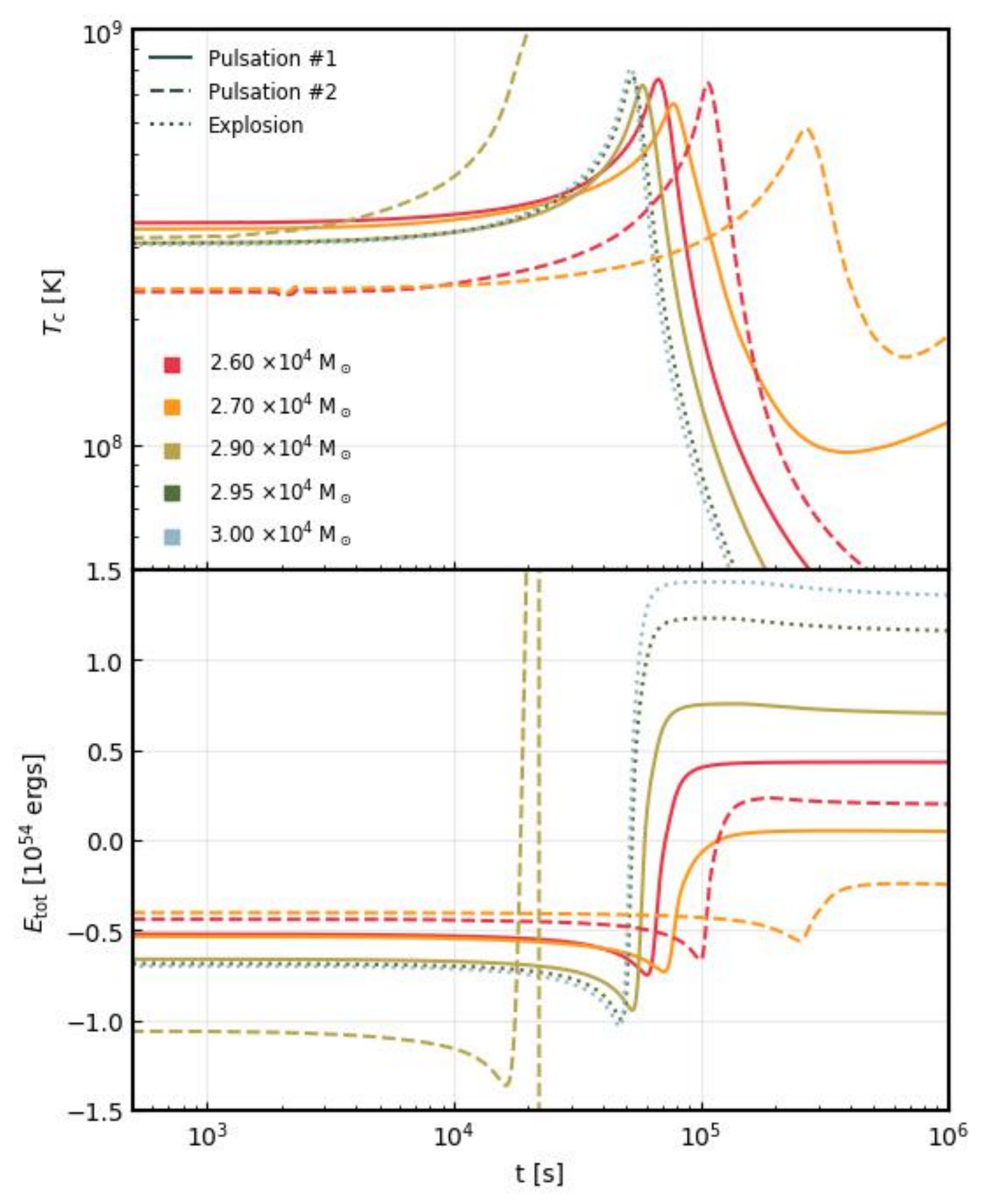}
    \caption{Time evolution in HYDnuc of central temperature (upper) and total energy (lower) for pulsation $\#1$ (solid lines) pulsation $\#2$ (dashed lines) and the explosions from \citet{Nagele2022arXiv220510493N} (dotted lines). Colors denote progenitor mass.}
    \label{fig:HYDnuc}
\end{figure}

At this point, we port the model profile back to HYDnuc using the same scheme as in \citet{Nagele2022arXiv220510493N}. HYDnuc is a 1D Lagrangian GR hydrodynamics code which includes all of the physics in HOSHI except for energy transfer. The model contracts rapidly in HYDnuc until alpha-process nuclear burning stalls the collapse and an outwards shock forms, which propagates to the surface of the star in $\sim 10^5$ s. Because we are interested primarily in the behavior of the core in HYDnuc, the surface resolution is poor, but this presents a challenge for accurately predicting the lightcurve of the supernova. In order to get around this difficulty, when we port the HYDnuc model to SNEC, we excise the outer meshes of the HYDnuc simulation where the radius intersects that of the HOSHI progenitor and replace the HYDnuc model outside of the corresponding mass coordinate with the HOSHI model. As the outermost region of the star has not changed during the HYDnuc simulation, using the higher resolution HOSHI profile will allow us to more accurately simulate shock breakout. We execute this procedure 100 steps before shock breakout in HYDnuc.

SNEC \citep{Morozova2015ascl.soft05033M} is an open source 1D Lagrangian hydrodynamics code which computes photosphere position and luminosity. From these we calculate effective temperature and assume a blackbody spectrum to determine AB magnitudes for the JWST filters. Our condition for porting to SNEC is that the explosion energy determined by SNEC must match that of HYDnuc to 1$\%$ accuracy, because we expect the explosion energy from HYDnuc to be accurate. In practice, this occurs once the shock has crossed into the region with small velocity ($\sim 10^{13}$ cm, Fig. 14 of \citealt{Nagele2022arXiv220510493N}) which is roughly determined by the sound speed multiplied by the explosion timescale. We follow the evolution of the explosion in SNEC until well past the end of the plateau phase identified in \citet{moriya2021}.

\section{Results}
\label{results}

\begin{table*}
	\centering
	\caption{Summary table for all models. The columns are total mass, outcome of HYDnuc, explosion energy, energy produced by nuclear reactions, ejecta mass, initial helium and oxygen fractions in the core, total (units of $\msun$) and percentage changes of helium and oxygen in the core.}
	\label{tab:models}
	\begin{tabular}{|c|l|r|r|r|r|r|r|r|r|r|} 
		\hline \hline
    		M [$10^4$ $\msun$] & Outcome & $E_{\rm exp}$ [ergs] & $E_{\rm nuc}$ [ergs] & $M_{\rm ej}$ [$\msun$] &$X_{\rm c}(^4$He)&$X_{\rm c}(^4$O) & $\Delta \; M(^4$He) & $\% \; M(^4$He)  &$\Delta \; M(^{16}$O) &$\% \; M(^{16}$O)   \\
    		\hline
2.60 & Pulsation $\#1$ & 4.32e53 & 6.95e53 & 2808 & 0.104 & 0.392 & 244.3 & 15.9 & 697.7 & 12.0 \\ 
 & Pulsation $\#2$ & 2.32e53 & 5.27e53 & 1175 & 0.090 & 0.345 & 180.4 & 12.3 & 489.5 & 10.7 \\ 
2.70 & Pulsation $\#1$ & 4.70e52 & 4.20e53 & 2299 & 0.222 & 0.479 & 149.1 & 4.5 & 443.3 & 6.2 \\ 
 & Pulsation $\#2$ & -2.42e53 & 1.23e53 & 0 & 0.215 & 0.448 & 43.5 & 1.4 & 129.0 & 2.1 \\ 
2.90 & Pulsation $\#1$ & 7.56e53 & 1.12e54 & 2078 & 0.589 & 0.221 & 360.2 & 3.8 & 806.6 & 22.5 \\ 
 & Collapse &  ---  &  ---  &  ---  & 0.566 & 0.172 &  ---  &  ---  &  ---  &  ---  \\ 
2.95 & Explosion & 1.23e54 & 1.49e54 & 29500 & 0.599 & 0.211 & 478.4 & 4.8 & 985.2 & 28.3 \\ 
3.00 & Explosion & 1.43e54 & 1.62e54 & 30000 & 0.652 & 0.165 & 519.0 & 4.7 & 955.5 & 34.4 \\ 
		\hline \hline
	\end{tabular}
\end{table*}

\begin{figure}
    \centering
    \includegraphics[width=\columnwidth]{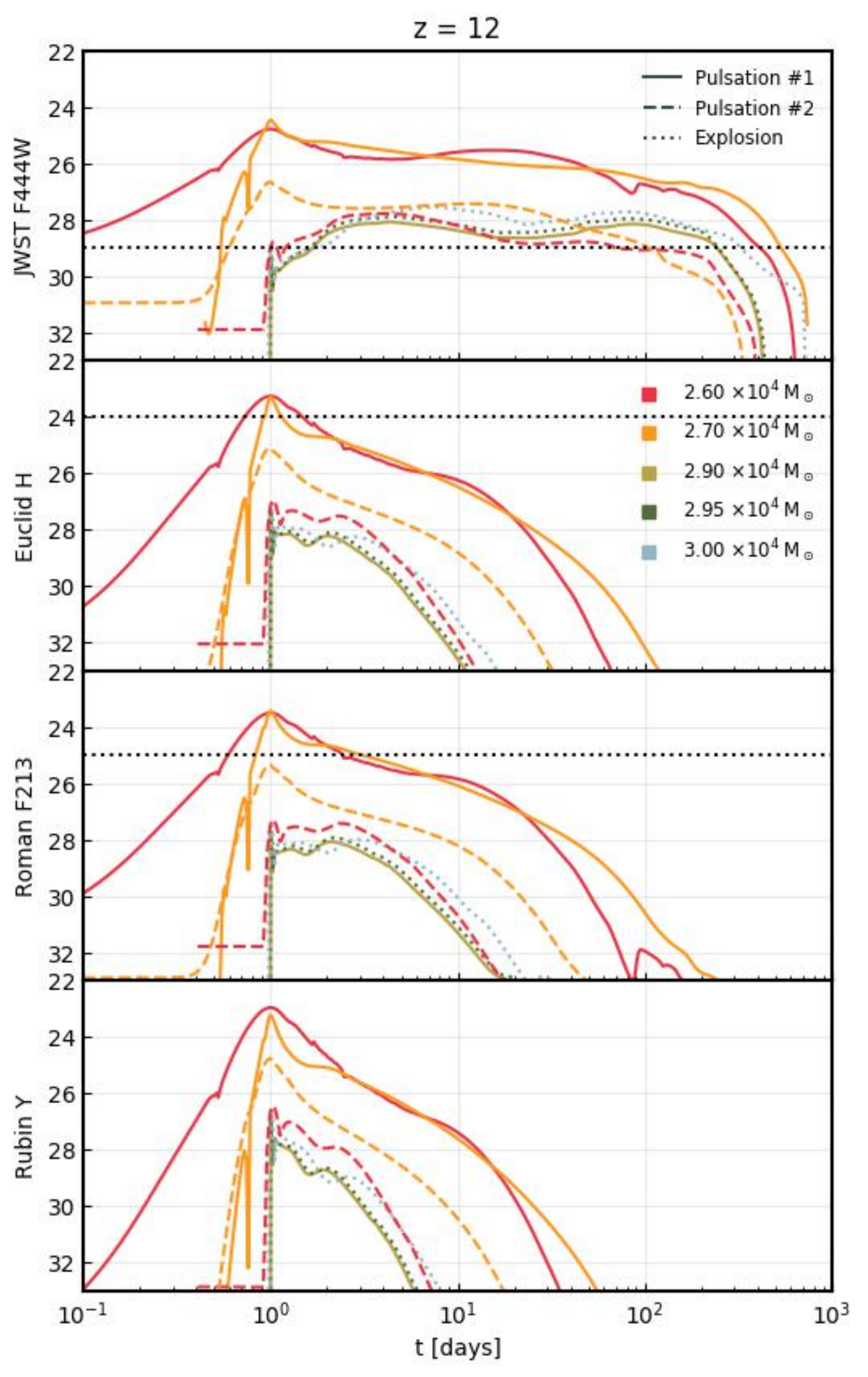}
    \caption{AB magnitudes at z = 12 for pulsation $\#1$ (solid lines) pulsation $\#2$ (dashed lines) and the explosions from \citet{Nagele2022arXiv220510493N} (dotted lines). The four panels show the reddest band of each of JWST, Euclid, Roman, and Rubin respectively. Horizontal dashed lines in the first three panels show typical limiting magnitudes of 29, 24, and 25 associated with each of these telescopes.}
    \label{fig:mag}
\end{figure}

\begin{figure}
    \centering
    \includegraphics[width=\columnwidth]{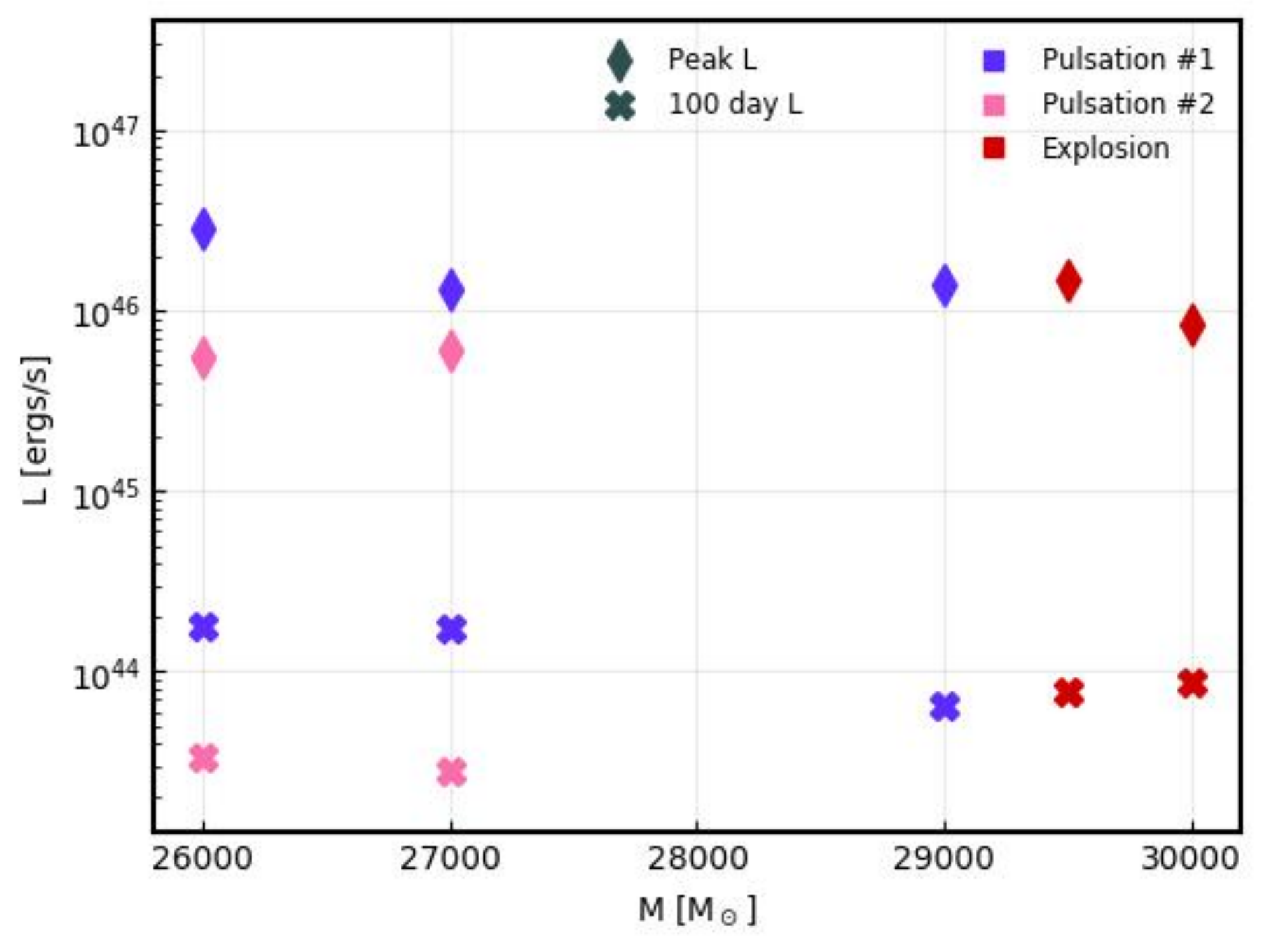}
    \caption{Peak luminosities (diamonds) and luminosities at 100 days (xs) shown as a function of progenitor mass (colors show type of event). }
    \label{fig:L}
\end{figure}

In Table \ref{tab:models}, we summarize the results of the pulsations together with the exploding models from \citet{Nagele2022arXiv220510493N}. The characteristics of the first pulsation have already been detailed so we will discuss the second pulsation for the three pulsating models in turn. The lowest mass model (26000 $\msun$) pulsates (at least) twice, with the second pulse having lower explosion energy and ejected mass than the first pulse. The energy produced via nuclear burning is also smaller in the second pulse, though the difference is moderate.

The intermediate mass model (27000 $\msun$) experiences a pulsation, but it is not strong enough to eject mass according to the criterion of \citet{Nagele2022arXiv220510493N}. This is because the maximum temperature is significantly lower than the other models (Fig. \ref{fig:HYDnuc}), resulting in less energy generation due to nuclear burning (Table \ref{tab:models}). We note that the final velocity structure suggests a small amount of mass ejection ($\sim 100$ $\msun$), and thus we will need to reexamine our mass ejection criterion in future work.

The highest mass model (29000 $\msun$) collapses to a black hole. However, we have confirmed that by artificially increasing the oxygen mass fraction by a small amount (0.17 to 0.2), we can induce an explosion which is a factor of three times more energetic than those found in \citet{Nagele2022arXiv220510493N}. Because the lower mass models have high oxygen mass fractions, it is likely that in between the intermediate mass model and the high mass model, there exists pulsating models which explode during the second pulse. We plan to investigate this phenomenon in future work.

Although we only analyze three models in this letter, hints of a trend are present. The lower and intermediate mass models have plenty of oxygen, but low levels of helium. In contrast, the higher mass model has plenty of helium, but low oxygen. This pattern suggests that somewhere in between these models, there may be a favorable regime for strong pulsations, where the stars have large reservoirs of both oxygen and helium, and this is another motivation for improving our mass resolution in a future study. Note that the oxygen reservoir is necessary because $^{12}$C($\alpha,\gamma$)$^{16}$O is subdominant in this temperature regime \citep[see ][]{Nagele2022arXiv220510493N} so that the explosive alpha process begins with $^{16}$O($\alpha,\gamma$)$^{20}$Ne.

The requirements for multiple pulsations are the following. First, the star must be stable during the contraction phase. after the first pulsation, so that the core can have time to remix. We have assumed this mixing in our study because the convection time-scale is much shorter than the contraction time-scale, and the convection criterion is satisfied in HOSHI; however, if for some reason this criterion were not satisfied, then the central region of the star might not contain the requisite fuel for the second pulsation. Next, the GR instability must occur soon after the end of the contracting phase. If this condition were not satisfied, then nuclear burning on evolutionary timescales would eventually deplete the helium reservoir required for the second pulsation. Finally, the helium and oxygen mass fractions after remixing cannot be too low. This last requirement is not satisfied by the 29000 $\msun$ model and is the reason for its collapse.

The stellar radius varies strongly between the different models. In particular, the lower mass models (26000, 27000) have large radii ($\sim 10^{15}$ cm) for the first pulse and smaller radii ($\sim 10^{14}$ cm) for the second pulse. The higher mass models also have relatively small radii. The photosphere is initially located at the stellar surface, but increases rapidly after shock breakout before stalling at $\sim 10^{16}$ cm due to hydrogen recombination. Because of the large radius of the photosphere, the effective temperature only briefly surpasses $10000$ K, and thus we have estimated that the amount of ionizing radiation is low and we do not expect this event to produce enough to effect the evolution of the proto-galaxy. However, our estimate is conservative, since the interaction between the ejecta and CSM (which might be interaction between the ejecta of successive pulsations) might produce additional and strong high-energy photons. We plan to investigate this possibility in future work.

The peak luminosity occurs soon after shock breakout, and is determined primarily by this radius which means that the pulsations from the low mass models radiate very strongly, despite their small explosion energies. For the first day (rest frame), the luminosity decreases monotonically as does the effective temperature. At around one day (or around four days for 26000 first pulsation), helium ions begin recombining and this leads to a slight uptick in the luminosity. Once the He ionization fraction has dropped, the luminosity continues to decrease until around twenty days, when hydrogen ions begin to recombine (visible in Fig \ref{fig:mag}). This is the beginning of the plateau phased identified in \citet{moriya2021}. The luminosity increases until the photosphere recession velocity matches the hydrodynamic expansion, at which point the photosphere stalls. 

In our models, the plateau phase is shorter than was found by \citet{moriya2021} by a factor of two (Fig \ref{fig:mag}). One possible explanation for the discrepancy is that the chemical distribution of our envelopes are different and hydrogen recombination terminates earlier. Another explanation could be that the inclusion of radiative transfer (and energy loss from the photo-sphere) is necessary to see the full duration of the plateau. For the GRSNe, (e.g. 30000 $\msun$), the luminosity in the plateau phase is also lower than in \citet{moriya2021} by a factor of 2 (Fig. \ref{fig:L}). This is likely due to the lower explosion energy in our models, though we note that \citet{moriya2021} initiated their light-curve calculation from the 2D model of \citet{chen2014} which had a higher explosion energy than the corresponding 1D model by roughly a factor of 2. Since we used a 1D model, the energy difference is a possible explanation for the discrepancy in plateau phase luminosity. At this point, we reiterate that even though the pulsations have smaller energies than the explosions (Table \ref{tab:models}), they have uniformly higher peak luminosities (with the exception of 26000 pulsation $\#2$) and the lower mass pulsations also have more luminous plateau phases (Fig \ref{fig:L}). This is due primarily to the larger radii of their progenitors.

\begin{figure}
    \centering
    \includegraphics[width=\columnwidth]{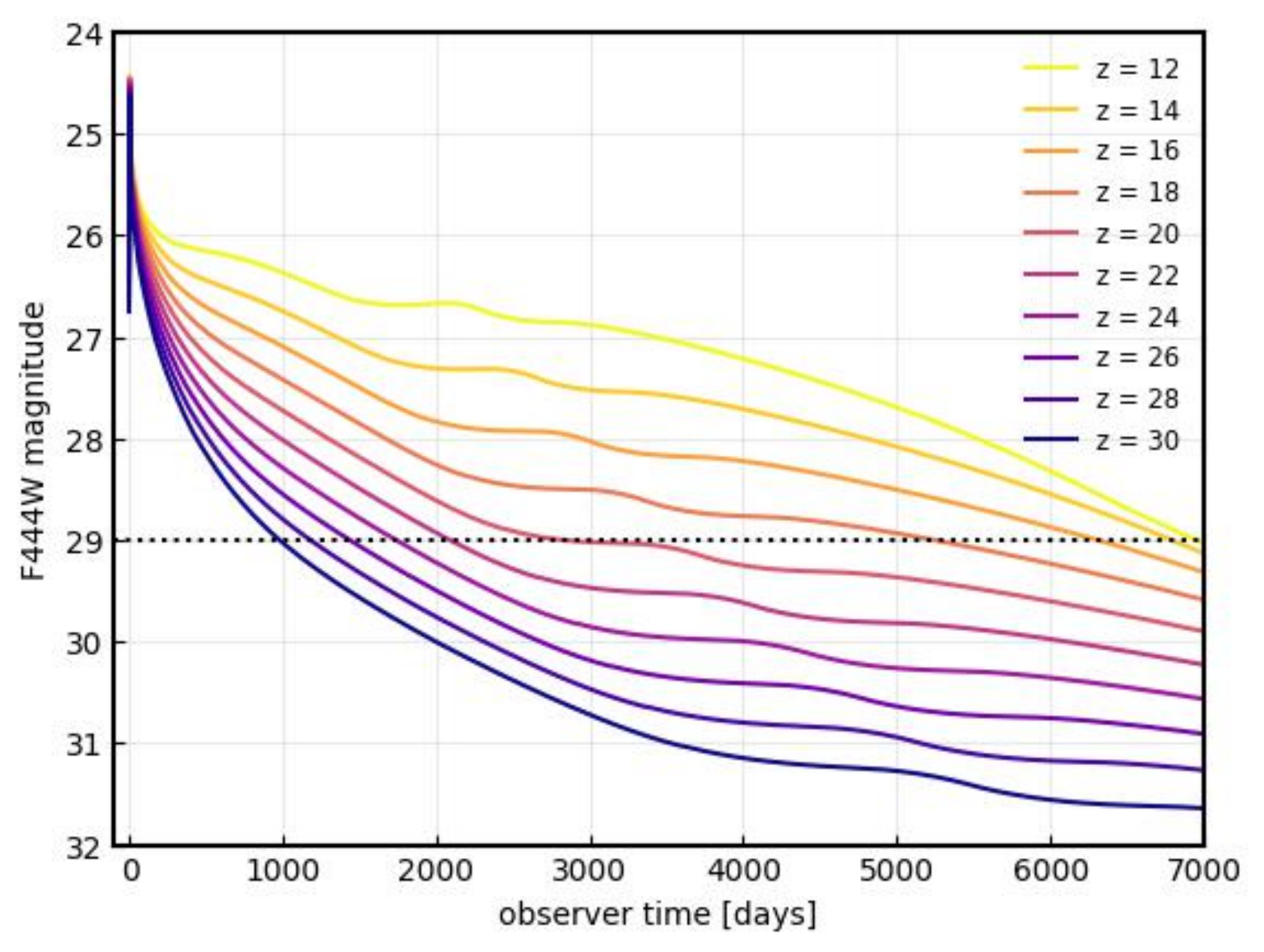}
    \caption{Redshift dependence of the F444W magnitude for our most luminous model, 27000 first pulsation. The x axis is shown in the observer frame. }
    \label{fig:z}
\end{figure}

We calculate the AB magnitude of our models in the reddest band of each of JWST, Euclid, Roman, and Rubin (Fig. \ref{fig:mag}). As in \citet{moriya2021}, we assume $\Lambda$CDM cosmology with $H_0 = 70$ km s$^{-1}$ Mpc$^{-1}$, $\Omega_M = 0.3$. JWST bands and cosmological distances were both calculated using astropy. We find that the plateau phase (100 days rest frame) of the second pulsation for 26000 $\msun$ is visible to JWST at z $>12$. Also of interest is plateau phase of the first pulsation, which is visible up to z $\sim 18$. At $z=30$, the first pulsation of 27000 $\msun$ is visible for more than 1000 days in the observer frame (Fig. \ref{fig:z}). Because of the long duration of these events, observing variations in the lightcurve might not always be feasible (especially during the plateau phase), but \citet{moriya2021} showed that multi-band observations can differentiate GRSNe from other persistent sources. The brighter first pulsations will also be visible to both Euclid and Roman at redshift 12 and lower, although the plateau phase will only be visible in the infrared. On the whole, our results may be slightly optimistic because we have not accounted for absorption or extinction within the interstellar or intra-cluster environment. In the future, we intend to do a more rigorous calculation of the JWST light curve. 

\section{Discussion}
\label{discussion}

The existence of supermassive stars and their contribution to the solution of the early universe supermassive black hole problem is unknown. If SMSs did exist, then the observation of the plateau phase of the GRSN by JWST or other future IR surveys represents one of our best chances at direct detection. Previously, we showed that the GRSN mass range was both lower and wider than initially thought \citep{Nagele2022arXiv220510493N}. In this paper, we have shown that as well as GRSNe, there exist pulsating models which effectively widens the mass range even further. Not only are these pulsations observable, but they are even brighter than the supernovae because of their extremely large radii. We have also shown that some models will experience multiple pulsations, each of which are potentially observable, although we should point out that in our current models, the second pulsation is less luminous than the first.

We will now now give a \textit{very} basic estimation of the event rate of GRSNe and pulsations. The formation scenario which has been investigated in the most detail is the radiation induced SMS formation scenario, and we base our estimates on the results of those investigations. The number density of DCBHs which result in SMSs with mass greater than 10000 $\msun$ has been estimated to lie anywhere in the range $10^{-1} - 10^{-9}$ cMpc$^{-3}$ \citep[][]{Woods2019PASA...36...27W}. Taking the very optimistic $10^{-2}$ cMpc$^{-3}$ \citep[][]{Agarwal2012MNRAS.425.2854A,Habouzit2016MNRAS.463..529H} results in tens of billions of SMSs between redshifts 7 and 30, which spans a time period of just over half a billion years. The empirically derived mass dependence of SMSs in \citet{Toyouchi2022arXiv220614459T}, i.e. $M^{-2.8}$ (Fig. 11) suggests that only $1.4\%$ of SMSs fall within the mass range of the explosions and pulsations investigated in this letter, so that 500 million GRSNe will have went off in this redshift range and volume. This works out to a rate of roughly one GRSN per year in the rest frame, or about one per decade in the observer frame.

Given the observing duration of one year in the rest frame, there is a reasonable chance that a GRSN is going off somewhere in the Universe at any given time. Unfortunately, JWST will not observe the whole sky, and the GRSNe will not be visible for a full year to wider surveys (Fig. \ref{fig:mag}), so an element of luck may be required for the detection of a GRSN with present and near future telescopes. The most promising avenues for altering this picture are repeated pulsations, longer observing duration due to CSM interaction, or different formation scenarios (e.g. baryonic dark matter streaming --- \citet{hirano2017}, turbulent cold flows --- \citet{Latif2022Natur.607...48L}, or galaxy mergers --- \citealt{Mayer2010Natur.466.1082M}) with higher density, higher metallicity thresholds, or more top heavy mass functions. The situation would also be improved by a deep infrared survey telescope which could observe the plateau phase at high redshift. All this being said, it is clear that a low density of SMSs, such as one comparable to the density of high redshift quasars, would be all but prohibitive to the detection of GRSNe, even with much more advanced telescopes.

The existence of the pulsations in \citet{Nagele2022arXiv220510493N} and of the multiple pulsations in this letter both require a detailed GR stability analysis. Specifically, a study using the Newtonian stability analysis will find the presence of neither pulsation (Fig. \ref{fig:HOSHI}). The same is true of a more naive GR stability analysis, such as following the evolution of a PN stellar evolution code \citep{chen2014,nagele2020}. We have not tried using the criterion of \citet{haemmerle2020} as it is only a sufficient condition for instability, but it is reasonable to assume that their criterion would find pulsations, because of the similarity in mass range (see Fig. 8 of \citealt{Nagele2022arXiv220510493N}).

Besides the GR stability analysis, the other ingredients to producing multiple pulsations are relativistic hydrodynamics and energy production via nuclear burning. We model these using established numerical methods, so all of the usual caveats apply. We have previously demonstrated the robustness of our parameter choices (Fig. 6 of \citealt{Nagele2022arXiv220510493N}). The potential spanner in the works is stellar rotation. In this letter, we have assumed the evolutionary models are extremely slow rotators, but in reality, they probably evolve closer to the $\Omega \Gamma$ limit \citep{hammerle2018}. Although this would still mean they are slow rotators, even a modest rotation can affect the evolution, and more critically, the GR stability \citep{Haemmerle2021A&A...650A.204H}, though much is still unknown about the interplay of rotation and GR. 

In the future, we plan to investigate two questions related to these pulsations. The first is how many times will these models pulse? For this, it is required that we update the mass ejection criterion from \citet{Nagele2022arXiv220510493N} because our criterion may have mischaracterized some pulsating models as stable ones. The second question is will these pulsations become interacting supernovae, or more specifically, what is the effect of circum-stellar material on the pulsation light curves? Circum-stellar material could originate from the SMS formation process or from previous pulsations.

\section*{Acknowledgements}

This study was supported in part by the Grant-in-Aid for the Scientific Research of Japan Society for the Promotion of Science (JSPS, Nos. JP19K03837, JP20H01905, JP20H00158, JP21H01123, JP20H00174, JP18H05223, JP22K20377) and by Grant-in-Aid for Scientific Research on Innovative areas (JP17H06357, JP17H06365) from the Ministry of Education, Culture, Sports, Science and Technology (MEXT), Japan. For providing high performance computing resources, YITP, Kyoto University is acknowledged.

\section*{Data Availability}

The data underlying this article will be shared on reasonable request to the corresponding author.




\bibliographystyle{mnras}
\bibliography{bib}




\bsp	
\label{lastpage}
\end{document}